\documentclass{iosart2x}

\usepackage{graphicx}
\usepackage{multirow}
\usepackage{multicol}
\usepackage{algorithm}
\usepackage{algcompatible}
\usepackage{amsmath}
\usepackage{algorithm}
\usepackage{booktabs}
\usepackage[noend]{algpseudocode}
\usepackage{xcolor}

\hypersetup{
    bookmarks    = false,            
    unicode      = false,            
    pdftoolbar   = false,             
    pdfmenubar   = true,             
}

\usepackage{cleveref}

\usepackage{amssymb}
\usepackage[width=1\textwidth]{caption}
\usepackage{array, bigdelim, makecell}  
\usepackage{comment}
\usepackage{xcolor}
\usepackage{enumitem}
\usepackage[labelformat=simple]{subcaption}

\pubyear{0000}
\volume{0}
\firstpage{1}
\lastpage{1}

\begin{document}
\makeatletter
\let\put@numberlines@box\relax
\makeatother

\begin{frontmatter}

\title{COV-ELM Classifier: An Extreme Learning Machine based identification of COVID-19 using Chest X-Ray Images}
\runtitle{COV-ELM Classifier: ELM based identification of COVID-19 using CXR Images}


\author[A]{\fnms{Sheetal} \snm{Rajpal}\ead[label=e1]{sheetal.rajpal.09@gmail.com}}
\author[B]{\fnms{Manoj} \snm{Agarwal}\ead[label=e2]{agar.manoj@gmail.com}\thanks{Corresponding Author}}
\author[A]{\fnms{Ankit} \snm{Rajpal}\ead[label=e3]{arajpal@cs.du.ac.in}}
\author[C]{\fnms{Navin} \snm{Lakhyani}\ead[label=e4]{navinlakhyani@gmail.com}}
\author[A]{\fnms{Arpita} \snm{Saggar}\ead[label=e5]{arpitasaggar.mca19.du@gmail.com}}
\author[A]{\fnms{Naveen} \snm{Kumar}\ead[label=e6]{nk.cs.du@gmail.com}}
\address[A]{Department of Computer Science, \orgname{University of Delhi},
Delhi, \cny{India}\printead[presep={\\}]{e1,e3,e5,e6}}
\address[B]{Department of Computer Science, Hansraj College, \orgname{University of Delhi},
Delhi, \cny{India}\printead[presep={\\}]{e2}}
\address[C]{Department of Radiology, \orgname{Saral Diagnostics, Pitam Pura},
Delhi, \cny{India}\printead[presep={\\}]{e4}}

\begin{abstract}
Coronaviruses constitute a family of viruses that gives rise to respiratory diseases. COVID-19 is an infectious disease caused by a newly discovered coronavirus also termed Severe acute respiratory syndrome coronavirus 2 (SARS-CoV-2). As COVID-19 is highly contagious, early diagnosis of COVID-19 is crucial for an effective treatment strategy. However, the reverse transcription-polymerase chain reaction (RT-PCR) test which is considered to be a gold standard in the diagnosis of COVID-19 suffers from a high false-negative rate. Therefore, the research community is exploring alternative diagnostic mechanisms. Chest X-ray (CXR) image analysis has emerged as a feasible and effective diagnostic technique towards this objective. In this work, we propose the COVID-19 classification problem as a three-class classification problem to distinguish between COVID-19, normal, and pneumonia classes. We propose a three-stage framework, named COV-ELM based on extreme learning machine (ELM). Our dataset comprises CXR images in a frontal view, namely Posteroanterior (PA) and Erect anteroposterior (AP). Stage one deals with preprocessing and transformation while stage two deals with feature extraction. These extracted features are passed as an input to the ELM at the third stage, resulting in the identification of COVID-19. The choice of ELM in this work has been motivated by its faster convergence, better generalization capability, and shorter training time in comparison to the conventional gradient-based learning algorithms. As bigger and diverse datasets become available, ELM can be quickly retrained as compared to its gradient-based competitor models. We use 10-fold cross-validation to evaluate the results of COV-ELM. The proposed model achieved a macro average F1-score of 0.95 and the overall sensitivity of ${0.94 \pm 0.02}$ at a 95\% confidence interval. When compared to state-of-the-art machine learning algorithms, the COV-ELM is found to outperform its competitors in this three-class classification scenario. Further, LIME has been integrated with the proposed COV-ELM model to generate annotated CXR images. The annotations are based on the superpixels that have contributed to distinguish between the different classes. It was observed that the superpixels correspond to the regions of the human lungs that are clinically observed in COVID-19 and Pneumonia cases.
\end{abstract}

\begin{keyword}
\kwd{COVID-19}
\kwd{Extreme Learning Machine}
\kwd{chest X-rays}
\kwd{Pneumonia viral}
\kwd{Pneumonia bacterial}
\end{keyword}

\end{frontmatter}








\section{Introduction}

Coronavirus disease 2019 (COVID-19), known to originate from Wuhan City in Hubei Province, China is a contagious infection resulting in respiratory illness in most cases. COVID-19 is caused by a novel coronavirus, widely recognized as severe acute respiratory syndrome coronavirus 2 (SARS-CoV-2; previously known as 2019-nCoV) \cite{ng2020imaging}. As the COVID-19 outbreak has become a global health emergency, on March 11, 2020, the WHO declared COVID-19 a global pandemic \cite{Archived20:online}. Moreover, COVID-19 disease shares similar characteristics as observed in other forms of viral or bacterial Pneumonia, making it difficult to separate between the two classes at the early stages. Thus, early accurate diagnosis of COVID-19 is critically important to contain the spread and the treatment of the affected subjects.

The reverse transcription-polymerase chain reaction (RT-PCR) test is popularly used for the detection of SARS-CoV-2. Although COVID-19 may be asymptotic in several instances, it has been reported that even many symptomatic cases showing characteristics of COVID-19 were not correctly diagnosed by RT-PCR test \cite{tahamtan2020real}. This has led to the search for alternative mechanisms that may be more accurate in the identification of COVID-19 disease. Traditionally, chest X-ray images (CXRs) have been the popular choice for diagnosis and treatment of respiratory disorders such as Pneumonia \cite{wang2017chestx,nanni2010local}. As a result, several research groups are working on developing models based on CXR images \cite{cohen2020covid,khan2020coronet,mahmud2020covxnet,wang2020covid}. However, most of them are struggling with the challenge to distinguish COVID-19 patients against those suffering from other forms of pneumonia \cite{khuzani2020covid}. 

Although deep neural networks have emerged as a popular tool for image-based analysis, these require tuning millions of parameters and search for the optimal value of hyper-parameters \cite{khan2020coronet, das2020automated, SHORFUZZAMAN2020107700, pathak2020deep, ch13,rajpal2021using}. Also, it is well known that the training of a deep neural network is a time-consuming task even on high-performance computing platforms.


Khan et al. \cite{khan2020coronet} proposed a deep convolutional neural network (DCNN) model to automate the detection of COVID-19 based on chest X-ray images. The model is based on Xception architecture \cite{chollet2017xception} pre-trained on ImageNet \cite{krizhevsky2012imagenet} and achieved an overall accuracy of 89.6\%.  Jain et al. \cite{jain2020deep} proposed a deep residual network for the automatic detection of COVID-19 in CXR image by differentiating it with the CXR images of bacterial pneumonia, viral pneumonia, and normal cases and exhibited an accuracy of 93.01\% in differentiating three classes using their first-stage model. They have further analyzed the CXR images showing the viral pneumonia features for the identification of COVID-19 case in their second stage model showing an exceptional performance with an accuracy of 97.22\%. Altan et al. \cite{altan2020recognition} used an efficient hybrid model consisting of two-dimensional (2D) curvelet transformation for the feature extraction, chaotic salp swarm algorithm (CSSA) to optimize the feature matrix, and EfficientNet-B0 model for the identification of COVID-19 cases. The model achieved an accuracy of 99.69\%. Mahmud et al. \cite{mahmud2020covxnet} proposed a DCNN model using a variation in dilation rate to extract distinguishing features from chest X-ray images and achieved an accuracy of 90.2\% for multi-class classification (COVID-19/Normal/Pneumonia). They also used Gradient-weighted Class Activation Mapping (Grad-CAM) to visualize the abnormal regions in CXR scans. Wang et al. \cite{wang2020covid} developed a computer-aided screening tool for detection of COVID-19 from CXR images based on a pre-trained network on ImageNet, tuned with the Adam optimizer, and achieved 91\% sensitivity for the COVID-19 class. Basu et al. \cite{basu2020deep} used domain extension transfer learning (DETL) framework comprising 12 layers. They used an already-trained network on the National Institutes of Health (NIH) CXR image dataset \cite{wang2017chestx} (comprising 108,948 frontal view X-ray images of 32,717 unique patients) which was fine-tuned for the COVID-19 dataset to obtain an overall accuracy ${95.3\% \pm 0.02}$ on 5-fold cross-validation. Marques et al. \cite{marques2020automated} made a novel attempt of applying EfficientNet \cite{tan2019efficientnet} (claimed to achieve an accuracy of 84.3\% top-1 accuracy on ImageNet) and evaluated their model using 10-fold stratified cross-validation method. 1092 samples have been used for training, and 122 images have been used for testing. They have achieved an average F1-score value of 0.97 in multi-class scenarios whereas 0.99 in the case of binary classification. Rajaraman et al. \cite{rajaraman2020iteratively} iteratively pruned the task-specific models (VGG-16, VGG-19, and Inception-V3) by pruning 2\% of the neurons in each convolutional layer and retrained the model to obtain a macro averaged F1-score of 0.99. Das et al. \cite{das2020automated} proposed a deep transfer learning approach for automated detection of COVID-19 disease. The network is fed with the features extracted using the Xception network. They obtained 97\% sensitivity for classifying COVID-19 cases from Pneumonia and respiratory diseases. They further show that their proposed model outperformed other popular deep networks such as VGGNet, ResNet50, AlexNet, GoogLeNet.  

Khuzani et al. \cite{khuzani2020covid} used multilayer neural networks (MLNN) to distinguish the CXR images of COVID-19 patients from other forms of pneumonia. They extracted a set of spatial and frequency domain features from X-ray images. Based on the evaluation of extracted features, they concluded that while Fast Fourier Transform (FFT) features were best suited in detecting the COVID-19, the normal class was best determined by the gray level difference method (GLDM). Principal Component Analysis (PCA) was applied to generate an optimized set of synthetic features that served as input to an  MLNN to distinguish COVID-19 images from the non-COVID-19 ones with an accuracy of 94\%. Rasheed et al. \citep{rasheed2021machine} applied PCA as a feature extraction technique resulting in 148 features. Further to investigate the suitability of the reduced feature set, CNN and logistic regression (LR) based models were developed to distinguish between COVID-19 and healthy cases using 250 CXR images belonging to each class. Accuracy of 100\% and 97.6\% for CNN and LR-based models respectively was reported.  

It is evident from the above discussion that so far the research groups have mainly focused on the use of deep neural networks which require millions of parameters and the optimal choice of hyper-parameters. However, it is well known that the training of a deep neural network is a time-consuming task even on high-performance computing platforms. Therefore, in order to improve the computational efficiency of the classification models, in this work, we have proposed the use of a single hidden layer feed-forward neural network (SLFN) known as extreme learning machine (ELM) \cite{huang2004extreme, akusok2015high}. The ELM is a batch learning algorithm proposed by Huang et al. \cite{huang2004extreme} and has been used extensively in different domains like ECG signal classification \cite{Karpagachelvi2012} and identification of arrhythmia disease \cite{Kim2009}. The ELM and its variants have also been applied in applications such as fingerprint identification \citep{Yang2013a}, lung cancer detection \citep{Daliri2012}, image and video watermarking \citep{rajpal2019novel,mishra2018bi}, and 3D object recognition \citep{Nian2013}. Govindarajan and Swaminathan \citep{govindarajan2021extreme} present a comparison of ELM and online-sequential ELM (OS-ELM) in the classification of tuberculosis from healthy subjects using CXR images. They have performed feature extraction using median robust extended local binary patterns and gradient local ternary patterns. ELM achieved a sensitivity value equal to 98.7\% while OS-ELM performed better with a sensitivity value of 99.3\%. Ismael and {\c{S}}eng{\"u}r \cite{ismael2020investigation} present ELM based binary classification model that uses multi-resolution approaches such as wavelet, shearlet, and contourlet transform for decomposition of CXR images. Features are extracted based on entropy and the normalized energy approaches. Using the ELM classifier, the sensitivity values obtained for wavelet, shearlet, and contourlet transforms are 96.07\%, 98.89\%, and 87.82\%, respectively. Thus, ELM is popularly applied in several domains due to its fast learning capability good generalization performance, and ease of implementation. 

The main contribution of this paper is to explore the suitability of ELM in the diagnosis of COVID-19 using CXR images. The faster convergence of ELM with only one tunable parameter made it more efficient as compared to conventional gradient-based learning algorithms. Another challenge addressed in this work is the identification of localized patterns to differentiate between the classes, namely, COVID-19, Pneumonia, and Normal. Further, to clinically establish the relevance of COV-ELM results, LIME has been integrated with it to generate annotated CXR images. These annotations represent regions that distinguish between the different classes.

The rest of the paper is organized as follows: \Cref{section3} gives the dataset description followed by the detailed methodology, preprocessing of the dataset, review of Extreme Learning Machine, outcomes of the experiments, and analysis of the results have been discussed in \Cref{section4}. Also, visualizations of COV-ELM results using LIME have been discussed in \Cref{LIME}. Finally, the conclusions and scope for future work are discussed in \Cref{section5}.

\section{Material and Methods}\label{section3}

\par In this section, we present a list of CXR image datasets used for experimentation in this work, followed by details of the proposed methodology. 

\subsection{Dataset Description}\label{dataDescription}

In the present work, we have used the following publicly available CXR datasets for COVID-19, Normal, and Pneumonia.
\begin{itemize}
\item COVID-19 Image Data Collection \cite{cohen2020covid}. It comprises 760 samples, COVID-19: 538, ARDS: 14, Other Diseases: 222.
\item COVID-19 Radiography Database (Kaggle) \cite{COVID19R73:online}. It comprises 2905 samples, COVID-19: 219, Normal: 1341, Viral Pneumonia: 1345.
\item Mendeley Chest X-ray Images \cite{kermany2018large}. It comprises 5856 samples, Pneumonia (Viral and Bacterial) : 4273, Normal:1583.
\end{itemize}

\begin{figure*}[!htbp]	
\centering
	\begin{subfigure}[t]{0.33\linewidth}
	\includegraphics[width=\linewidth]{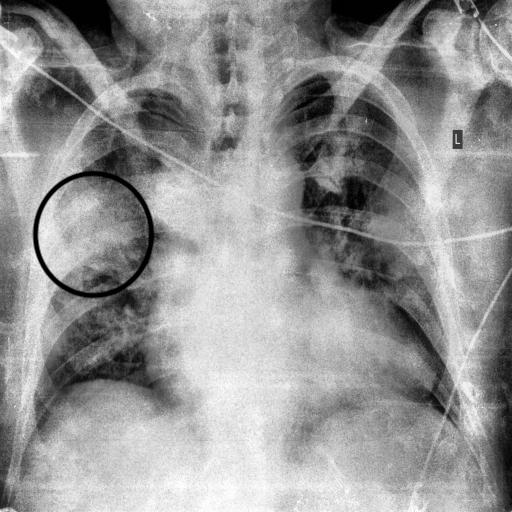}
	\caption{COVID-19}  \label{cxrcovid}
	\end{subfigure}
	\hfill
	\begin{subfigure}[t]{0.33\linewidth}
	\includegraphics[width=\linewidth]{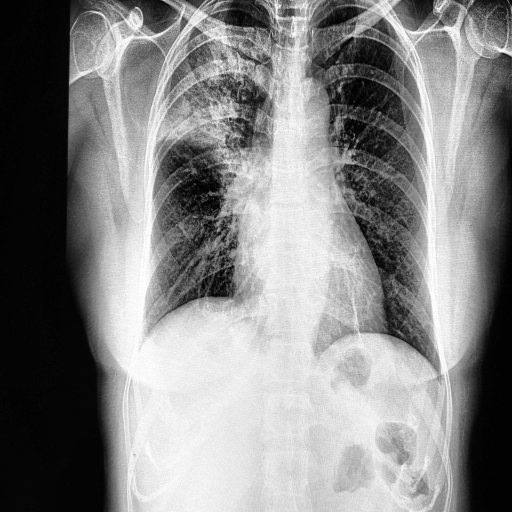}
	\caption{Pneumonia}  \label{cxrpneumonia}
	\end{subfigure}
	\begin{subfigure}[t]{0.33\linewidth}
	\includegraphics[width=\linewidth]{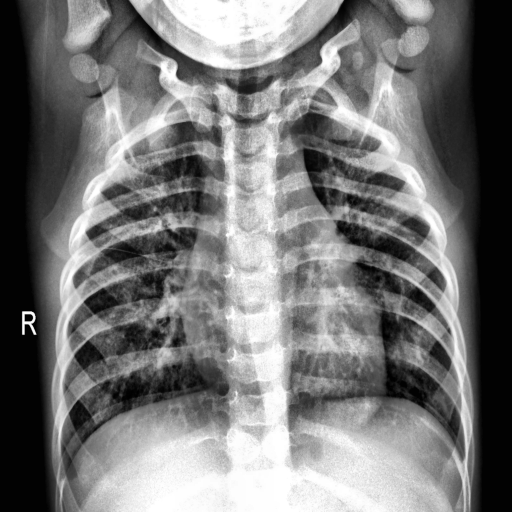}
	\caption{Normal}  \label{cxrnormal}
	\end{subfigure}
\caption{Manually annotated CXR images highlighting the regions of interest that distinguishes between COVID-19 and Pneumonia cases. The above regions are marked by a radiologist after clinical evaluation of these CXR images.}
\label{cxrimages}
\end{figure*}

In this work, we only consider the CXR images in a frontal view, namely Poster anterior (PA) and Erect anteroposterior (AP). The first two databases in the above list comprise 520 such images. For the training purpose, we have used these images along with 520 CXR images of normal and pneumonia cases from COVID-19 Radiography Database (Kaggle) \cite{COVID19R73:online} and Mendeley Chest X-ray Images \cite{kermany2018large}. \Cref{cxrcovid,cxrpneumonia} depicts the manually marked region of interest that distinguishes between COVID-19 and Pneumonia cases in CXR images. The above regions are marked by a radiologist after clinical evaluation of these CXR images.  

\subsection{Preprocessing}\label{preprocesDetail}

\begin{figure*}[!htbp]
\centerline{\includegraphics[width=6in]{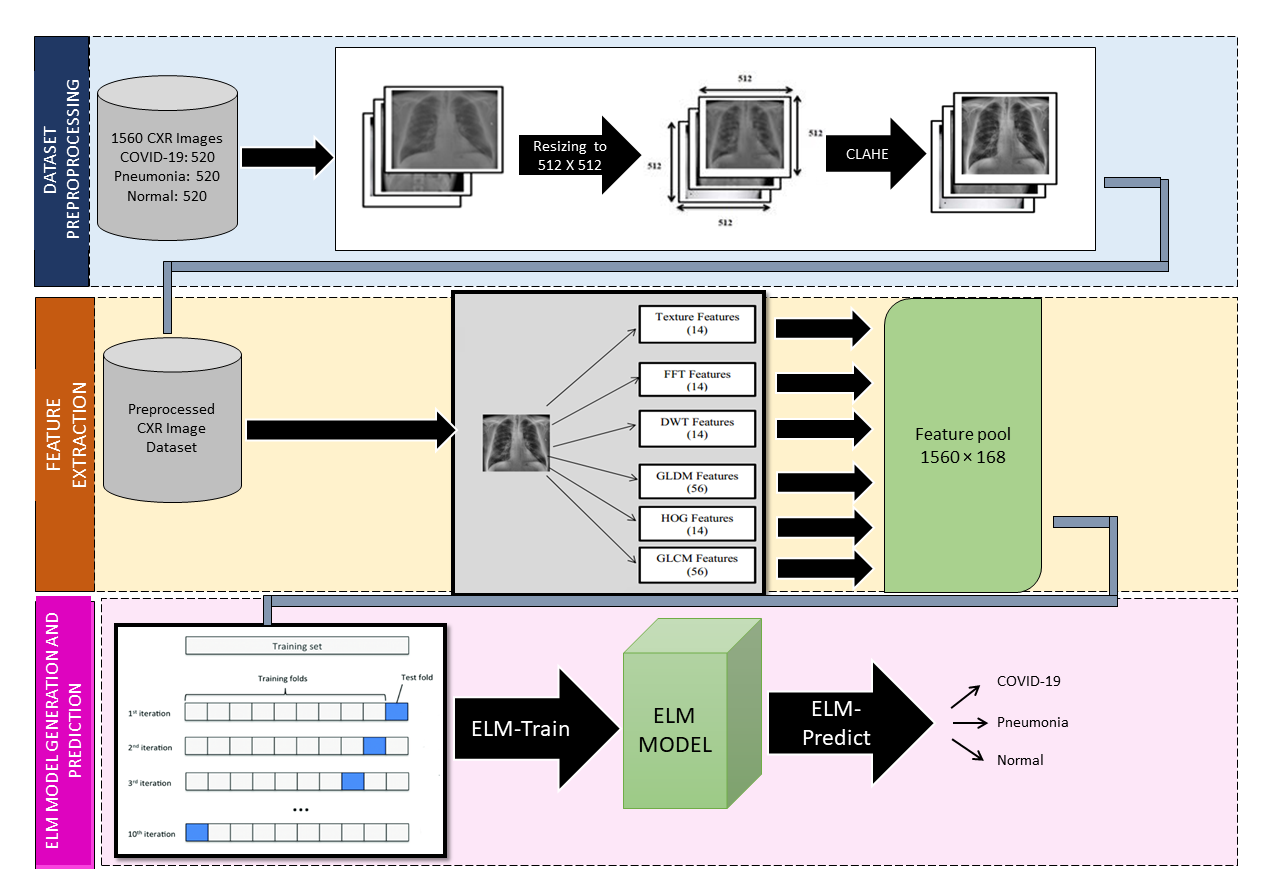}}
\caption{COV-ELM Framework: Dataset Preprocessing, Feature Extraction, and ELM based classification model.}
\label{GraphicalAbstract}
\end{figure*}

Due to diversity in the CXR image collection, they are resized and subjected to min-max normalization \cite{jain2005score} to ensure uniformity. Further, to enhance the local contrast in the CXRs, Contrast Limited Adaptive Histogram Equalization (CLAHE), a variant of adaptive histogram equalization is applied. \Cref{GraphicalAbstract} depicts the framework of the three-staged proposed model. In stage one, the preprocessing includes resizing, normalization, and  CLAHE \cite{ahmad2012analysis} applied in the sequence shown. The preprocessed CXRs are passed to stage 2 of the framework for feature extraction. 

\subsection{Feature Extraction}\label{featureExtract}

Texture plays a significant role in the identification of the region of interest (ROI) and classification of images \cite{haralick1973textural}. In this stage, we consider two types of features: texture and frequency-based as shown in \Cref{GraphicalAbstract}. The texture features consisted of four groups. The first group of features is directly generated from the preprocessed image of \begin{math} 512 \times 512 \end{math}. These include area, mean, standard deviation, skewness, kurtosis, energy, entropy, max, min, mean absolute deviation, median, range, root mean square, and uniformity. Remaining texture features are obtained by applying gray-level co-occurrence matrix (GLCM) \cite{zare2013automatic,TextureA25:online}, histogram of oriented gradients (HOG) \cite{dalal2005histograms, Histogra27:online, xue2015chest}, and gray-level difference matrix (GLDM) \cite{kim1999statistical, khuzani2020covid}. Apart from texture features, the use of frequency features also plays an important role in developing efficient classifiers in medical imaging \cite{shree2018identification,leibstein2006detecting,parveen2011detection}. In the present work, the frequency features are extracted using Fast Fourier Transform (FFT) and Discrete Wavelet Transform (DWT). Zargari et al. \cite{zargari2018prediction} used the aforementioned statistical features for predicting chemotherapy response in ovarian cancer patients. Drawing inspiration from their work, we computed these features for the FFT map and three-level (LL3) DWT  coefficients to generate a vector of frequency features. Finally,  a vector of features is obtained by concatenating the textural feature vector of length (140) with the frequency vector of length (28) to generate a vector of size \begin{math} 168 \end{math} for each CXR image.

\subsection{Extreme Learning Machine}\label{ELM}

\begin{figure}[!htbp]
\centerline{\includegraphics[width=\linewidth]{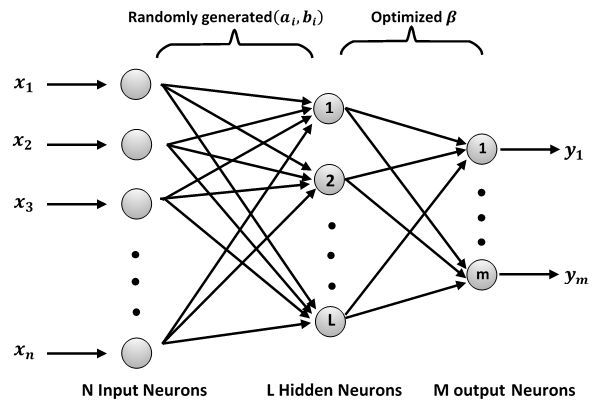}}
\caption{ELM Architecture: The ELM network comprises an input layer, a hidden layer, and an output layer}
\label{ELMarch}
\end{figure}

In stage three, the features extracted at stage 2 are passed as input to the Extreme Learning Machine (ELM) based classification model as shown in \Cref{GraphicalAbstract}. ELM was proposed by Huang et al. as an efficient alternative to the backpropagation algorithm for single-layer feed-forward networks (SNFN) \cite{huang2004extreme}. It is a fast learning algorithm with good generalization performance as compared to other traditional feed-forward networks. An ELM works by initializing a set of weights randomly and computing the output weights analytically by Moore-Penrose Matrix Inverse \citep{Fill2000}. \Cref{ELMarch} depicts the overall ELM architecture and the details of its functioning are provided in \Cref{elm_algo}. 

Given a training set \begin{math}(x_{j} , t_{j}), x_{j} \in \textbf{R}^{n} \end{math}, \begin{math} t_{j} \in \textbf{R}^{m}  \end{math} for \begin{math} j=1,2,\dots,N \end{math}, where the pairs \begin{math}(x_{j} , t_{j}) \end{math} denote the training vectors and the corresponding target values, following \cite{huang2004extreme},  the standard ELM having  \begin{math} L \end{math} nodes is modeled as:

\begin{equation}
\sum_{i=1}^{L}\beta_{i}g_{i}(a_{i}.x_{j}+b_{i})=t_{j}
\label{eq:elmmodel}
\end{equation}

In \Cref{eq:elmmodel}, \begin{math} a_{i} \end{math} denotes the weight vector that connects the input layer to the \begin{math} i^{th} \end{math} hidden node and \begin{math} b_{i} \end{math} denotes the corresponding bias. Further, \begin{math} \beta_{i} \end{math} denotes the weight vector connecting the \begin{math} i^{th} \end{math} hidden node and the output neurons. The above N equations may also be  represented as:

\begin{equation} G\beta=T \label{eq:G} \end{equation}

\noindent The form of the hidden layer output matrix \begin{math} G \end{math}, mentioned in \Cref{eq:G}, is given in \Cref{eq:Gmat}. The form of vectors \begin{math} \beta \end{math} and \begin{math} T \end{math} is given in \Cref{eq:betamat}. 

\begin{gather}
G=
\begin{bmatrix}
   g(a_{1}.x_{1}+b_{1}) & \dots & g(a_{L}.x_{1}+b_{L}) \\
   \vdots & \dots & \vdots \\
   g(a_{1}.x_{N}+b_{1}) & \dots & g(a_{L}.x_{N}+b_{L})
\end{bmatrix}_{N \times L}
\label{eq:Gmat}
\end{gather}

\begin{gather}
\beta=
\begin{bmatrix}
   \beta^{T}_{1} \\
   \vdots \\
   \beta^{T}_{L}
\end{bmatrix}_{L \times m}
and \quad 
T=
\begin{bmatrix}
   t^{T}_{1} \\
   \vdots \\
   t^{T}_{N}
\end{bmatrix}_{N \times m}
\label{eq:betamat}
\end{gather}

\noindent The solution of the above system of linear equations is obtained using Moore-Penrose generalized inverse (\Cref{eq:betacap}).

\begin{equation}
{\beta} = G^{\dagger}T
\label{eq:betacap}
\end{equation}

In \Cref{eq:betacap}, \begin{math}G^{\dagger}=(G^{T}G)^{-1}G^{T}\end{math} denotes the Moore-Penrose generalized inverse \citep{Fill2000} of matrix G.

\begin{algorithm}[H]
\caption{ELM Algorithm}\label{elm_algo}
\begin{itemize}[label={}]
\item \textbf{Input:}
\item \quad \quad Training set: \begin{math}(x_{j} , t_{j}), x_{j} \in \textbf{R}^{n} \end{math}, \begin{math} t_{j} \in \textbf{R}^{m}  \end{math} for \begin{math} j=1,2,\dots,N \end{math}
\item \quad \quad Activation function: \begin{math} $g$ \colon \textbf R \to \textbf R \end{math}
\item \quad \quad Number of hidden nodes: L
\item \textbf{Output:} 
\item \quad \quad Optimized weight matrix: \begin{math} \beta \end{math}
\begin{enumerate}
\item Randomly assign hidden node parameters \begin{math} (a_{i},b_{i}), i=1,2,\dots,L \end{math};
\item Compute the hidden-layer output matrix G;
\item Compute output weight vector \begin{math} {\beta} = G^{\dagger}T \end{math}
\end{enumerate}
\end{itemize} 
\end{algorithm}

Huang et al. \cite{Huang2006} argue that ELM outperforms the conventional learning algorithms in terms of learning speed, and in most of the cases shows better generalization capability than the conventional gradient-based learning algorithms such as backpropagation where the weights are adjusted with a non-linear relationship between the input and the output \cite{Fill2000}. They further stated that ELM can compute the desired weights of the network in a single step in comparison to classical methods.

\subsection{COV-ELM} \label{trainTestELM}

In this work, we use ELM discussed in \Cref{ELM} to develop an ELM classifier (COV-ELM) for the detection of COVID-19 in CXR images.  Based on experimentation, we used L2-normalized radial basis function (rbf-l2) activation function. We also experimented with the different number of neurons in the hidden layer. Using 10-fold cross-validation, we observed that with an increase in the number of neurons in the hidden layer, accuracy increases up to ${L=140}$  neurons, and the highest 10-fold cross-validation accuracy of 94.74\% was reached when the number of hidden neurons was ${L=350}$. Experimenting with different seeds, we found the peak accuracy was reached for the number of hidden neurons in the range 350 to 380 but without any further increase in 10-fold cross-validation accuracy. So, for further experiments, we fixed the number of hidden neurons as ${L=350}$. 

\begin{figure}[!htbp]
\centerline{\includegraphics[width=3in]{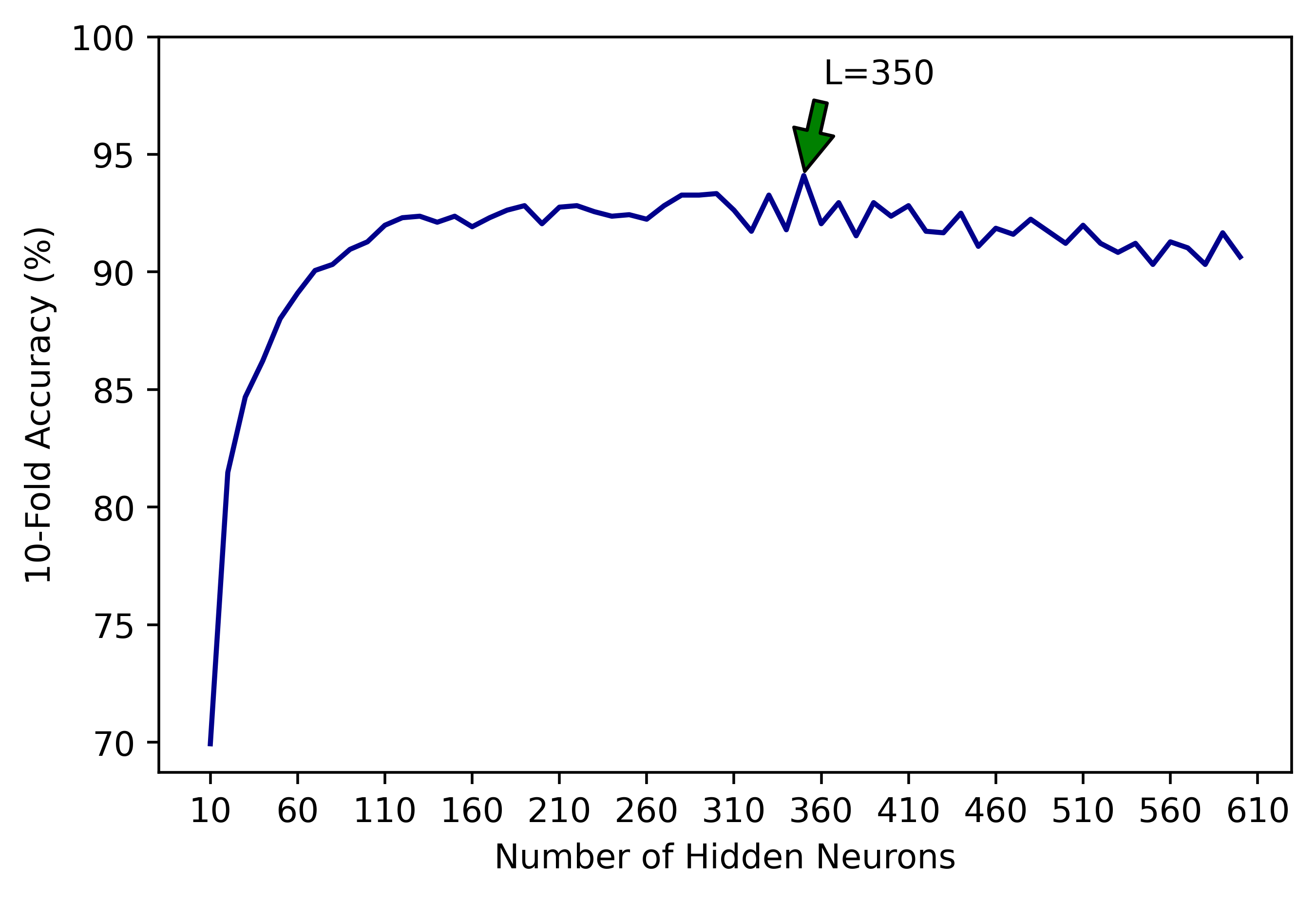}}
\caption{Effect of the increase in the number of hidden neurons (${L}$) on 10-fold cross-validation accuracy. Accuracy increases with increase in ${L}$ upto ${L=140}$, and witnessed highest 10-fold cross-validation accuracy of 94.74\% at ${L=350}$}
\label{LvsAccuracy}
\end{figure}

\begin{figure}[!htbp]
\centerline{\includegraphics[width=2.5in]{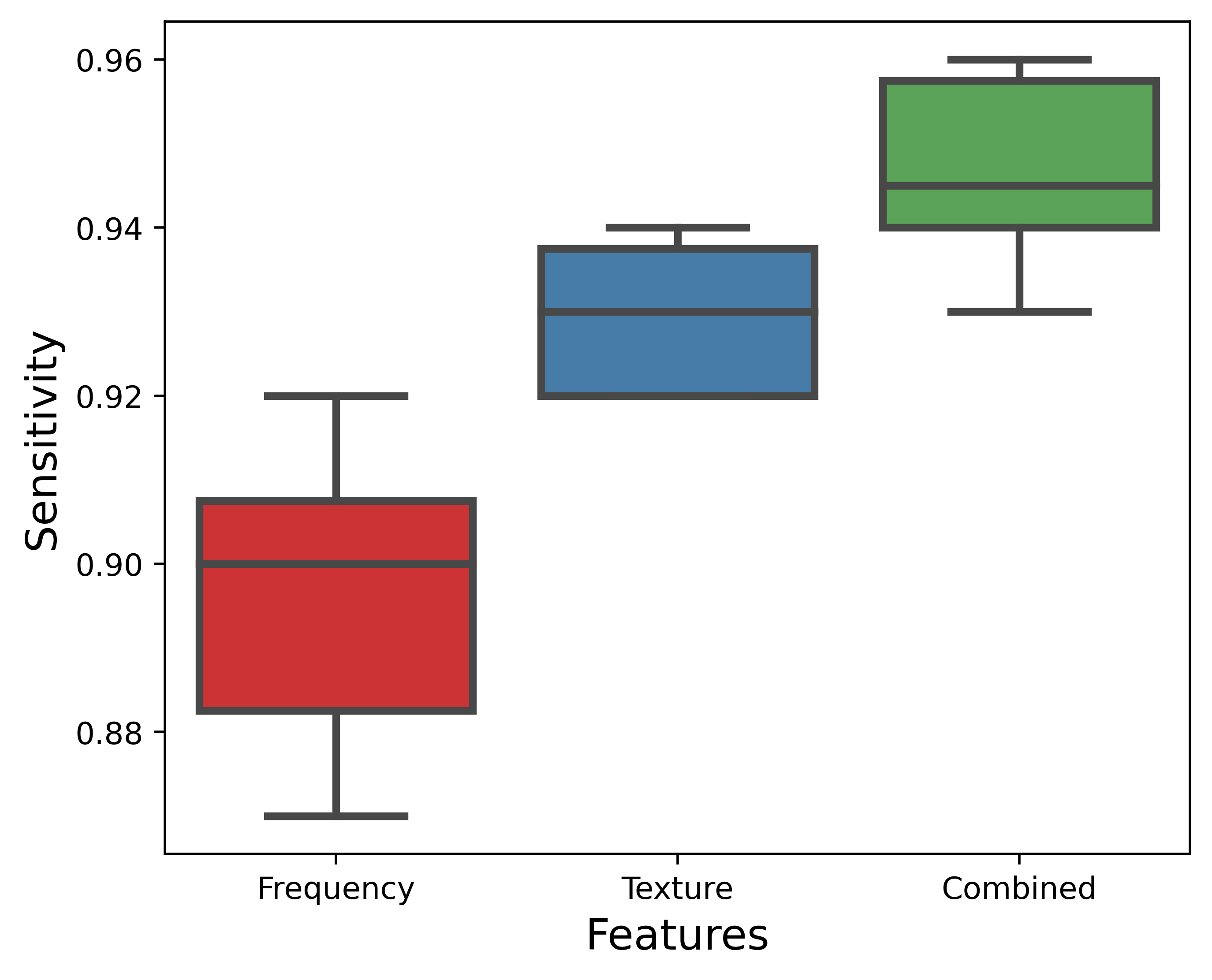}}
\caption{Boxplot for sensitivity (recall) values using frequency features, texture features, and combined set of frequency and texture features. The combined set of features depicts the median sensitivity of 0.945 which scores over the median values considering frequency and texture features separately.}
\label{boxplot}
\end{figure}

Boxplot in \Cref{boxplot} depicts the variation in sensitivity value. It is evident from the results that the texture features score over frequency features. We also examined the influence of a combined set of features ($168$) on the classification process. It may be noted that the model yields median sensitivity of 0.945 using the combined set of features which scores over the median sensitivity values considering the frequency and texture features separately, exhibiting 0.90 and 0.93 respectively.

\section{Results and Discussion} \label{section4}

\begin{figure*}[!htbp]	
\centering
	\begin{subfigure}[b]{0.48\linewidth}
	\includegraphics[width=\linewidth]{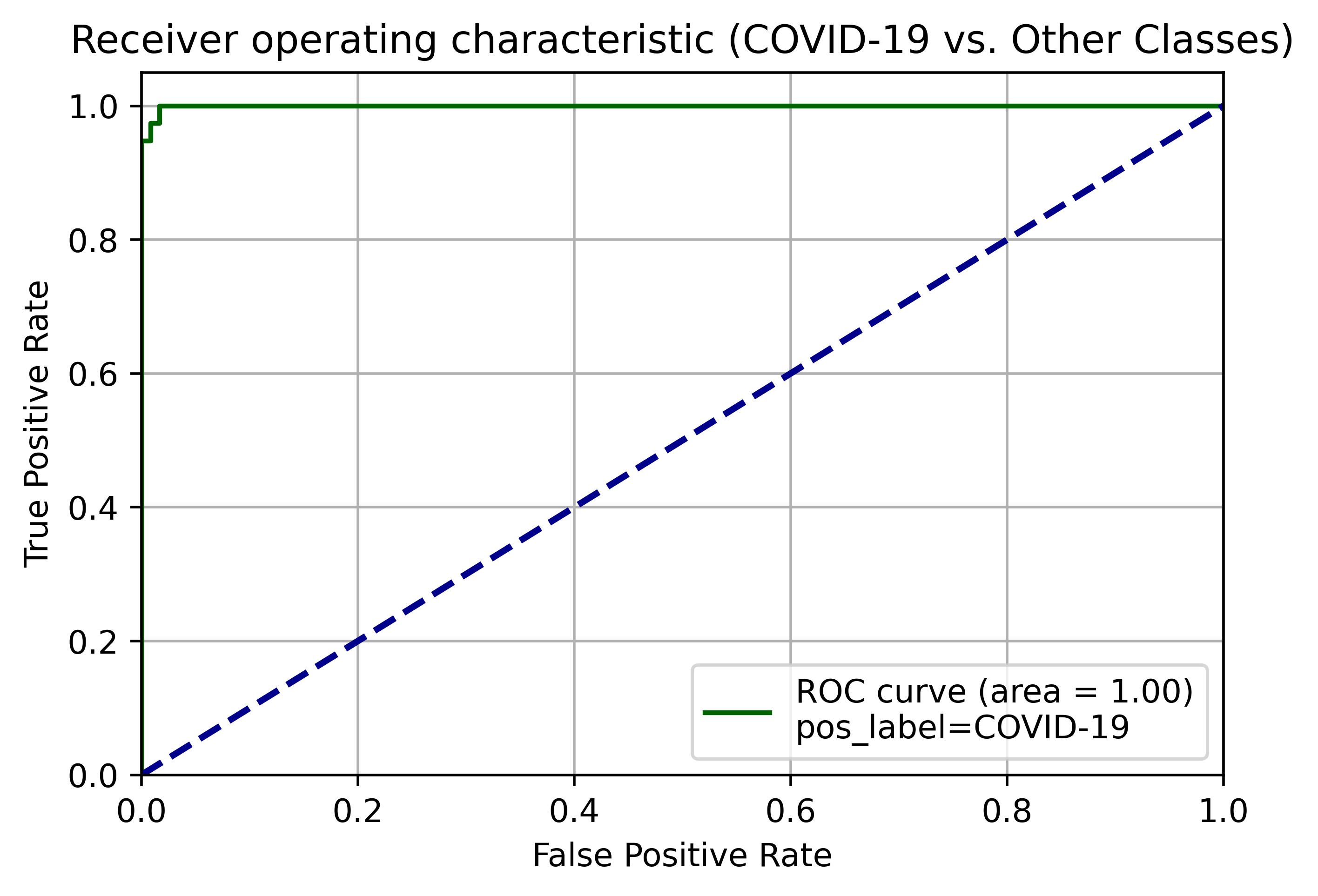}
	\caption{ROC Curve (COVID-19 vs other classes)} \label{roc-covid}
	\end{subfigure}
	\begin{subfigure}[b]{0.48\linewidth}
	\includegraphics[width=\linewidth]{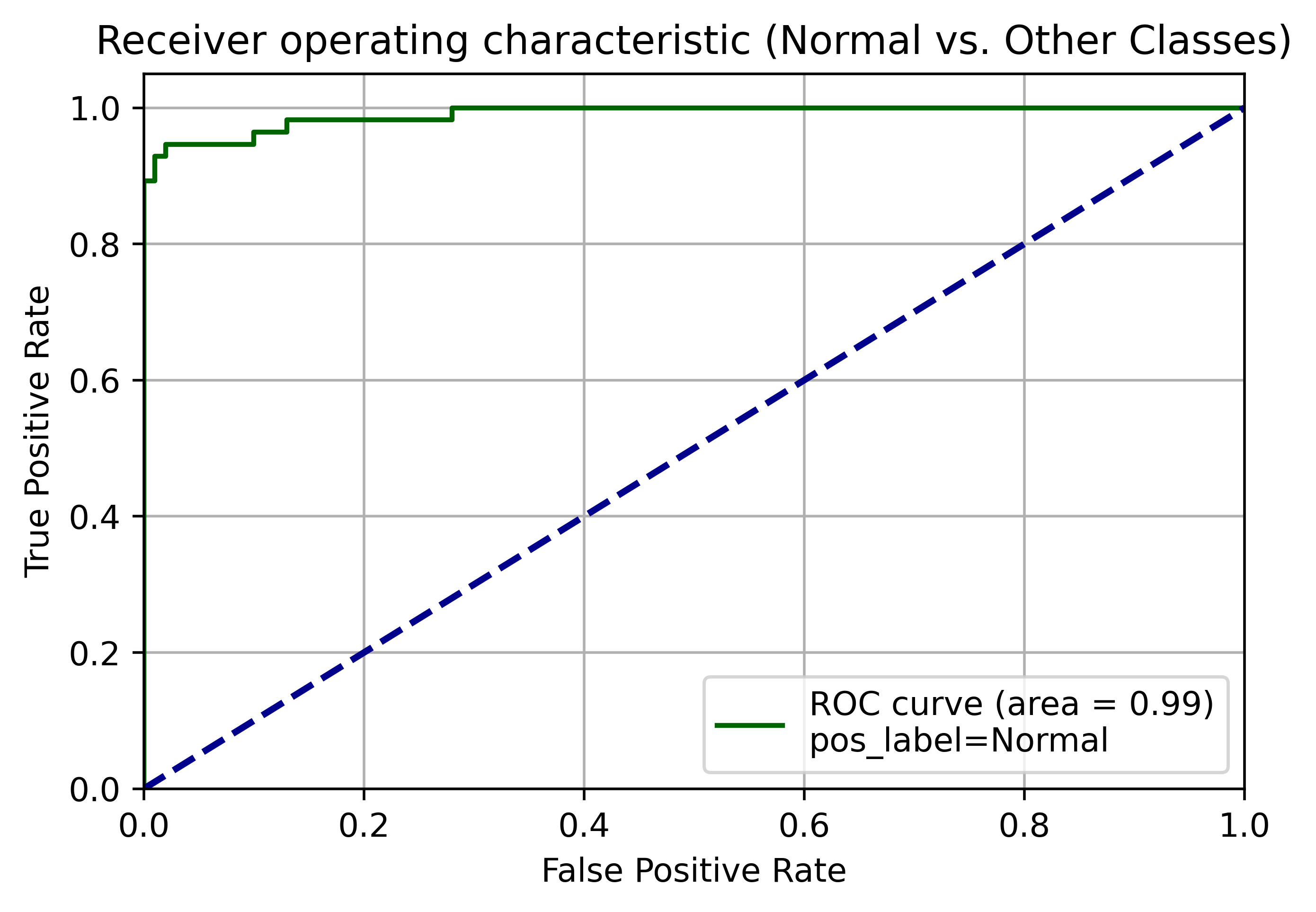}
	\caption{ROC Curve (Normal vs other classes)}  \label{roc-normal}
	\end{subfigure}
	\begin{subfigure}[b]{0.48\linewidth}
	\includegraphics[width=\linewidth]{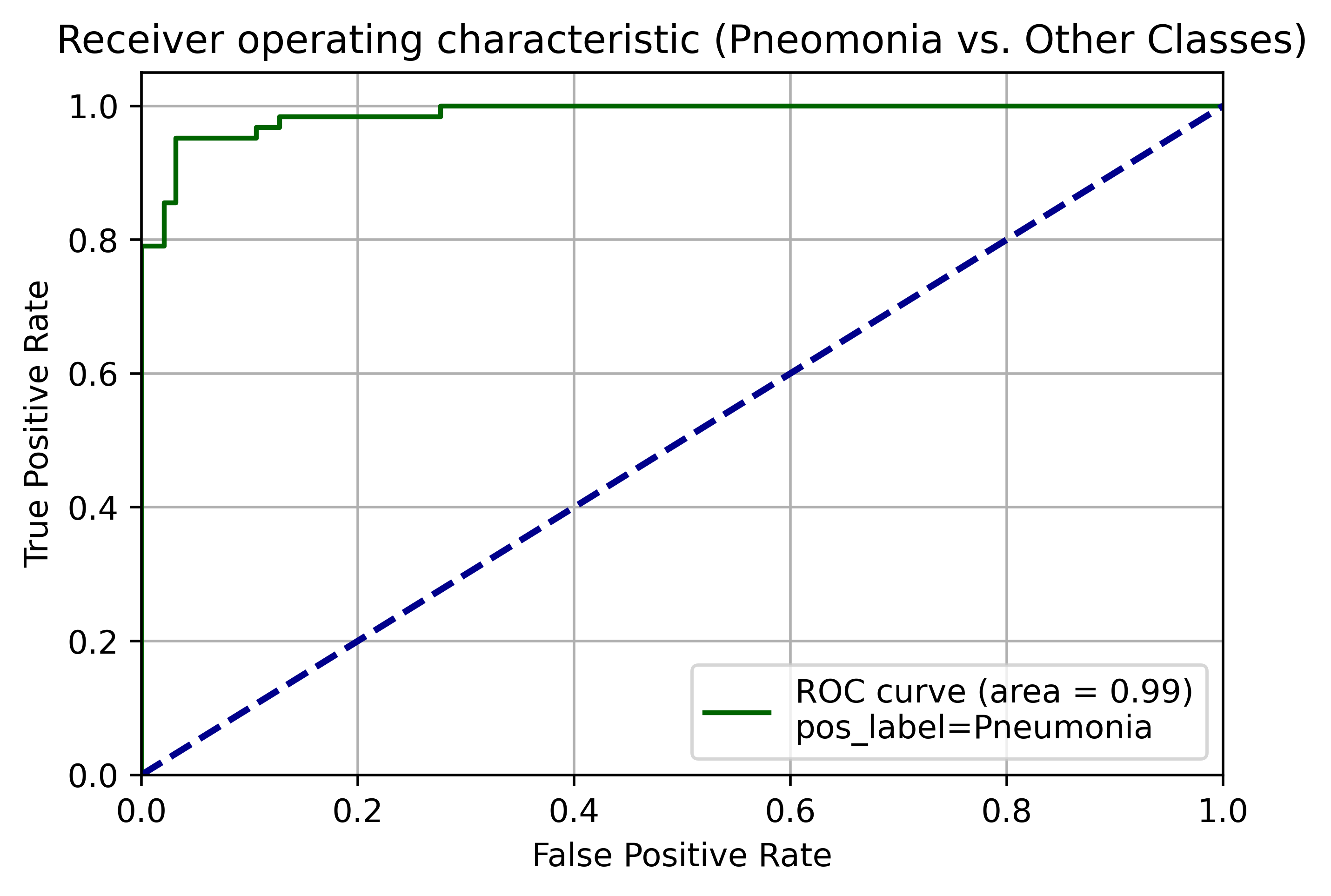}
	\caption{ROC Curve (Pneumonia vs other classes)}  \label{roc-pneumonia}
	\end{subfigure}
\caption{AUC is near unity for each of the three classes namely COVID-19, Normal, and Pneumonia in one vs all setting.}
\label{AUC}
\end{figure*}

We have carried out all the experiments using Python 3.6.9 on the NVIDIA Tesla K80 GPU provided by Google Colaboratory. To evaluate the performance of the proposed method for the three-class classification problem, we trained the model on the CXR dataset using 10-fold cross-validation. Following Handy and Till \cite{hand2001simple}, we depict the receiver operating characteristic (ROC) curves for each of the three classes, namely COVID-19, Normal, and Pneumonia for one fold (please see in \Cref{AUC}). It is apparent from the ROC curves that AUC is near unity for all three classes which shows a good generalization performance of COV-ELM.    

\begin{figure*}[!htbp]	
\centering
	\begin{subfigure}[t]{0.45\linewidth}
	\includegraphics[width=\linewidth, height=1.8in]{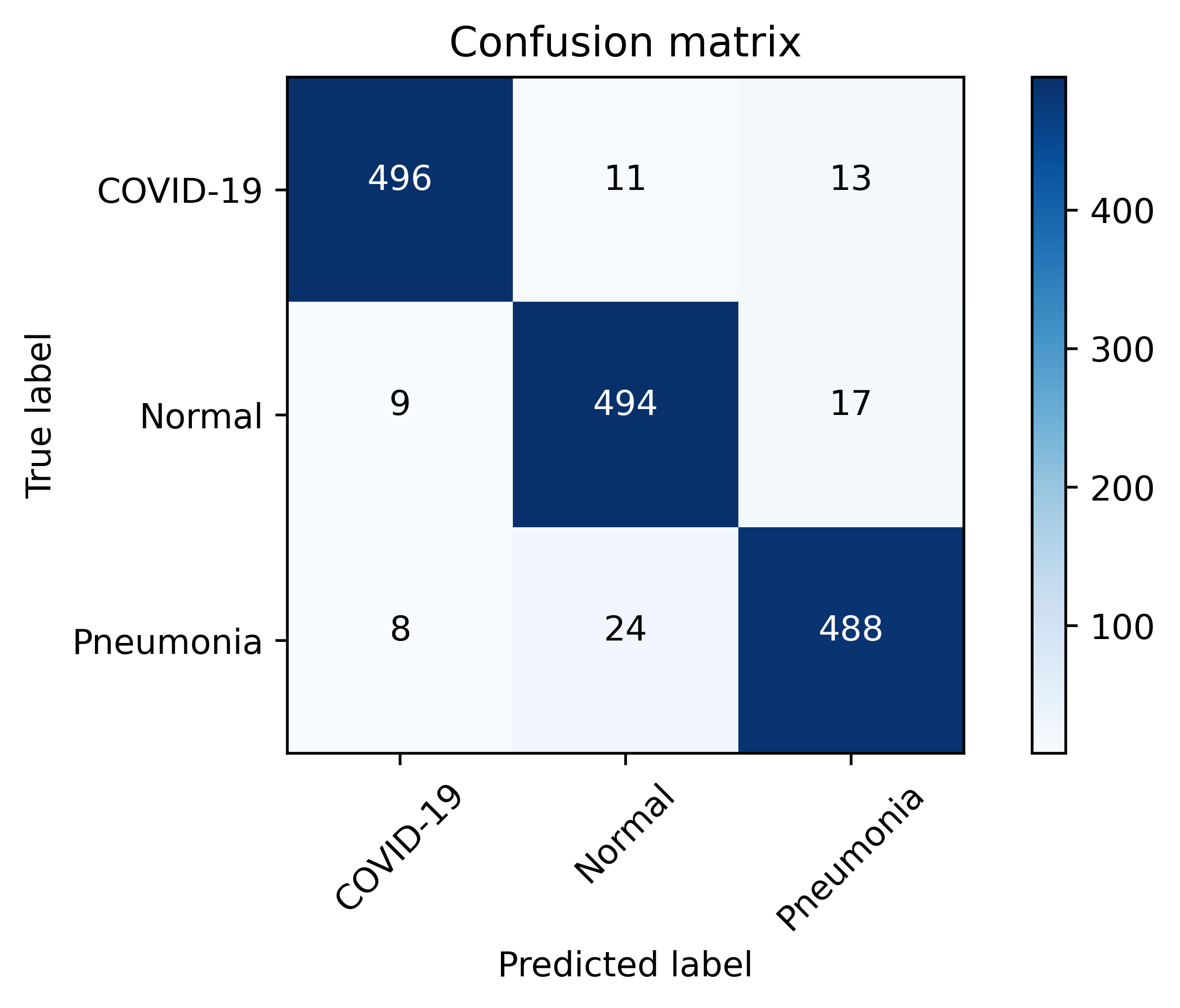}
	\caption{Confusion matrix summarizing number of instances correctly classified across diagonal elements.}  \label{confusion}
	\end{subfigure}
	\hfill
	\begin{subfigure}[t]{0.45\linewidth}
	\includegraphics[width=\linewidth, height=1.8in]{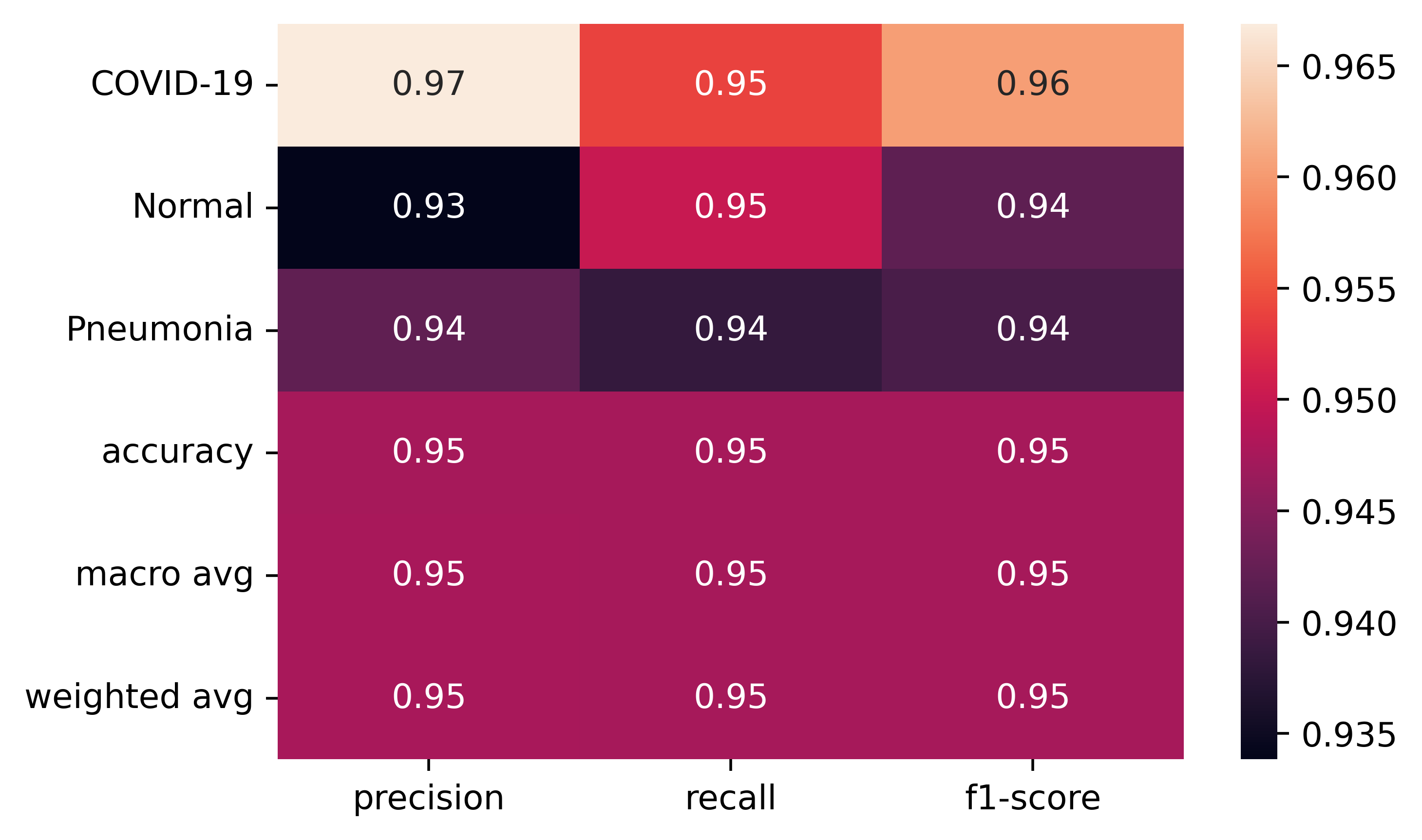}
	\caption{Heatmap summarizing performance metrics: precision, recall and f1-score}  \label{heatm}
	\end{subfigure}
\caption{The classification error in classifying COVID-19, Normal, and Pneumonia is 4.62\%, 5\%, and 6.16\% respectively and the macro average of f1-score is 0.95.}
\label{confuHeatmap}
\end{figure*}

To evaluate the performance of the proposed classifier, we carried out 10-fold cross-validation. \Cref{confuHeatmap} depicts the confusion matrix and the heatmap for 10-fold cross-validation. The results of the 10-fold cross-validation are summarized in a confusion matrix (\Cref{confusion}). It shows that out of 520 COVID-19 patients, 496 were correctly identified, eleven were misclassified as normal and thirteen were labeled as pneumonia. Similarly, pneumonia and normal subjects were also labeled by the system quite accurately. Thus, we obtained an overall accuracy of 94.74\% and a high recall rate of  95.38\%, 95.00\%, and 93.84\% for COVID-19, Normal, and Pneumonia classes respectively. The macro average of the f1-score is 0.95 as depicted in the heatmap (\Cref{heatm}).  As shown in \Cref{tableRecall}, COV-ELM identified COVID-19, Normal, and Pneumonia classes with sensitivity  ${0.95 \pm 0.04}$, ${0.95 \pm 0.01}$, and ${0.94 \pm 0.03}$ respectively at 95\% confidence interval.

\begin{table}[!htbp]
\centering
\caption {Sensitivity (Recall) values for COVID-19, Normal, and Pneumonia at 95\% confidence interval.} 
\label{tableRecall}
\begin{tabular}{|c|c|c|}
\hline
\multicolumn{3}{|c|}{\textbf{Sensitivity at 95\% CI}}             \\ \hline
COVID-19          & Normal            & Pneumonia         \\ \hline
${0.95 \pm 0.04}$ & ${0.95 \pm 0.01}$ & ${0.94 \pm 0.03}$ \\ \hline
\end{tabular}
\end{table}

To establish the effectiveness of our approach, the COV-ELM is compared with the state-of-the-art machine learning algorithms, namely  support vector classifier (SVC) using rbf and linear kernels, gradient boosting classifier (GBC), random forest ensemble (RBE), artificial neural networks (ANN), decision tree classifier (DTC), and voting classifier (VC) ensemble of (logistic regression (LR), SVC, and GBC) in terms of sensitivity at 95\% confidence interval (CI) (please see \Cref{tableComparison}). It is clear that COV-ELM has higher sensitivity as compared to its competitors. It is evident from the table that the proposed approach achieves a sensitivity of ${0.94 \pm 0.02}$ and accuracy of ${0.94 \pm 0.03}$ which scores over other state-of-the-art classifiers.

\begin{table}[!htbp]
\centering
\caption {Comparison of COV-ELM with other state-of-the-art classifiers in terms of sensitivity and accuracy values at 95\% confidence interval.} 
\label{tableComparison}
\begin{tabular}{|l|c|c|}
\hline
Classifier                                    & Sensitivity     &    Accuracy \\ \hline
ELM (L=350, rbf-l2)                           & ${0.94 \pm 0.02}$    &  ${0.94 \pm 0.03}$ \\
GBC (learning rate=1.0)                       & ${0.91 \pm 0.05}$   &  ${0.91 \pm 0.04}$   \\
SVC (C=1.0, kernel='rbf')                     & ${0.86 \pm 0.06}$   &  ${0.86 \pm 0.05}$   \\
SVC (C=1.0, kernel='linear')                  & ${0.90 \pm 0.05}$   & ${0.90 \pm 0.06}$  \\
RBE (min\_samples\_split=2)                   & ${0.89 \pm 0.05}$   & ${0.89 \pm 0.04}$  \\
ANN (23,747 Parameters)                       & ${0.85 \pm 0.08}$  & ${0.85 \pm 0.07}$ \\
DTC (min\_samples\_leaf=1)                    & ${0.82 \pm 0.07}$  &  ${0.82 \pm 0.06}$      \\
VC (LR, SVC, GBC) 														& ${0.89 \pm 0.05}$  &  ${0.89 \pm 0.06}$      \\ \hline
\end{tabular} 
\end{table}

Recently, Sayg{\i}l{\i} Ahmet \citep{saygili2021new} proposed the use of  machine learning techniques such as bag of tree, kernel ELM (K-ELM), k-nearest neighbor (k-NN), and SVC to detect COVID-19 cases using CXR images. \Cref{tableASC} shows a comparison between the aforementioned work \citep{saygili2021new} and the proposed approach (COV-ELM).

\begin{table}[!htbp]
\centering
\caption {Comparison of COV-ELM with the recently proposed approach by Sayg{\i}l{\i} Ahmet \citep{saygili2021new} for the detection of COVID-19 using CXR images.} 
\label{tableASC}
\resizebox{7cm}{!}{\begin{tabular}{|l|l|c|}
\hline
 \multicolumn{1}{|c|}{\textbf{Dataset Used}}                                                               & \multicolumn{1}{c|}{\textbf{Technique}} & \textbf{\begin{tabular}[c]{@{}c@{}}COVID-19\\      Sensitivity (\%)\end{tabular}} \\ \hline
\begin{tabular}[c]{@{}l@{}} \textbf{Proposed (COV-ELM)} \\ COVID-19:   520\\ Normal: 520\\ Pneumonia:520\end{tabular}    & ELM (L=350, rbf-l2)                                                                                                                    & \textbf{94.74}                                                                 \\ \hline

       & \begin{tabular}[c]{@{}l@{}} Bag of tree \\ (\# of trees=100)\end{tabular}                                                                 & 71.20                                               \\
 \begin{tabular}[c]{@{}l@{}} \textbf{Sayg{\i}l{\i} Ahmet \citep{saygili2021new}} \\ COVID-19:   125\\ Normal: 500\\ Pneumonia:500\end{tabular}        & \begin{tabular}[c]{@{}l@{}} K-ELM \\(L=4096, rbf, \\${C=1e-1}$) \end{tabular}                                                                & 88.00                                               \\
         & \begin{tabular}[c]{@{}l@{}} \\k-NN (k=1,\\ Minkowski distance) \end{tabular}                                                                & 94.40                                               \\
         & \begin{tabular}[c]{@{}l@{}} \\SVC (Default) \end{tabular}                                                                & 88.80                                               \\ \hline                 
                   
\end{tabular}}
\end{table}


\section{Visualization using LIME} \label{LIME}
In order to corroborate the COV-ELM results with clinical findings, we have used a recently proposed AI tool -- Local Interpretable Model-agnostic Explanations (LIME) \cite{ribeiro2016should}. LIME perturbs an input image and helps in analyzing the effect of these perturbations on the predictions of a given machine learning model. 

\begin{figure*}[!htbp]
\centering
	\begin{subfigure}[t]{\linewidth}
	\includegraphics[width=\linewidth]{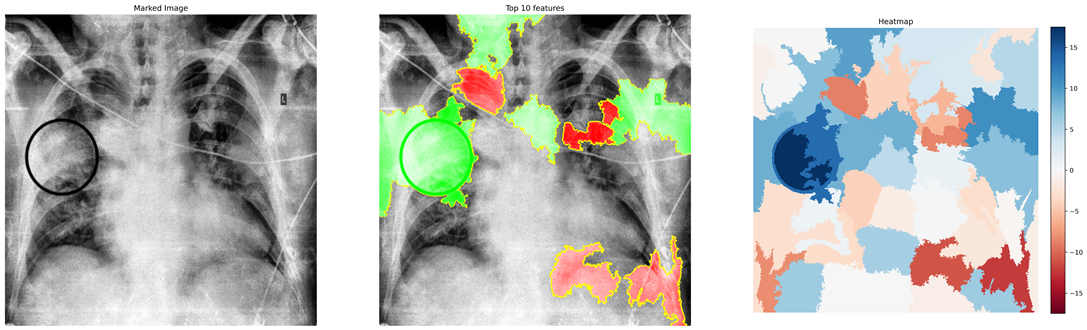}
	\caption{COVID-19}  \label{limecxrcovid}
	\end{subfigure}
	\hfill
	\begin{subfigure}[t]{\linewidth}
	\includegraphics[width=\linewidth]{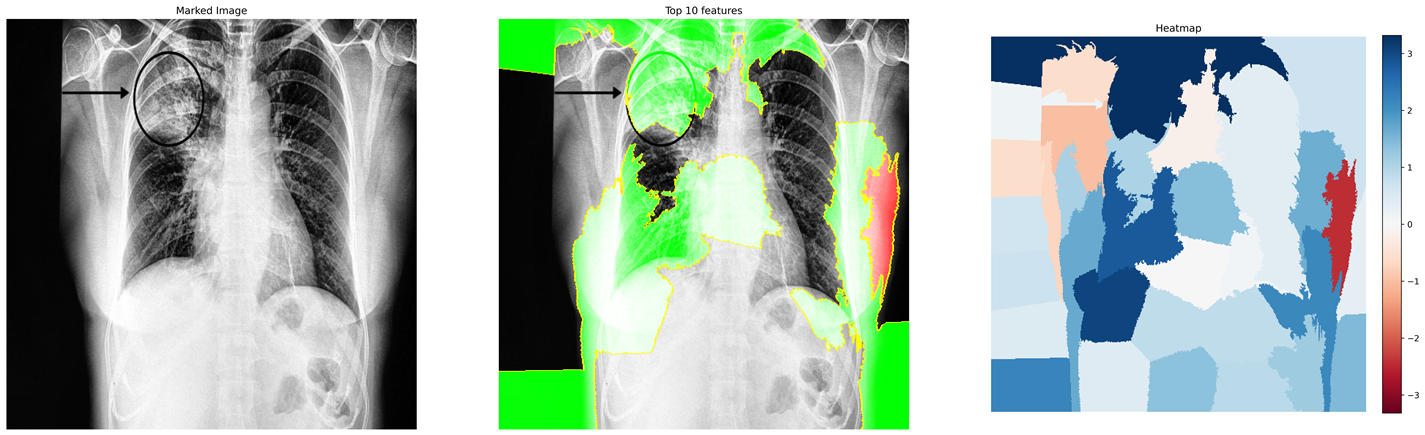}
	\caption{Pneumonia}  \label{limecxrpneumonia}
	\end{subfigure}
	\begin{subfigure}[t]{\linewidth}
	\includegraphics[width=\linewidth]{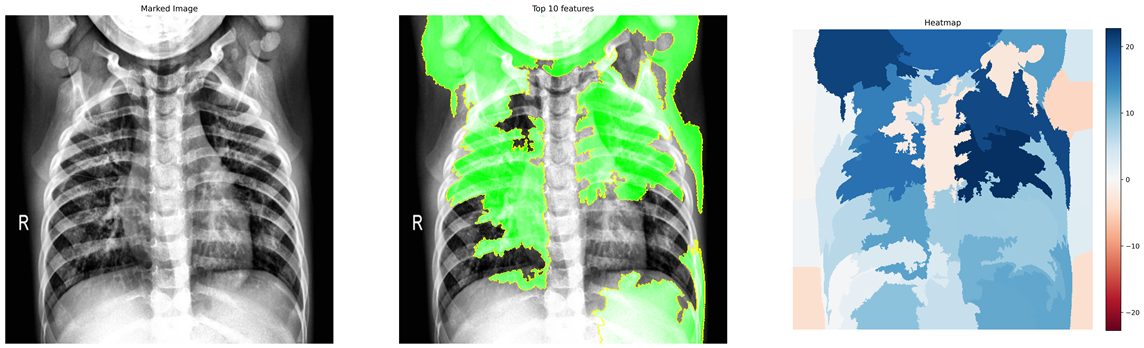}
	\caption{Normal}  \label{limecxrnormal}
	\end{subfigure}
\caption{(a)-(c) corresponding to COVID-19, Pneumonia, normal cases, respectively. In each row -- the first CXR image depicts the clinically evaluated and manually marked regions, second CXR image highlights the top 10 superpixels obtained using LIME, and the third image is the LIME generated heatmap corresponding to the second image.}
\label{limecxrimages}
\end{figure*}

\Cref{limecxrimages} (a)-(c) shows images relating to COVID-19, Pneumonia, normal cases, respectively. Each subfigure in a row comprises three images of the same patient relating to a medical condition. In each row, the clinical condition has been marked by a radiologist in the first image. In the second image in the same row, the top 10 superpixels obtained using LIME have been marked using green and red colors. Superpixels contributing toward and against the predicted class appear in green and red colors, respectively. Finally, the third image in the same row depicts the LIME-generated heatmap corresponding to the second image. The intensity of the blue color of a particular region in the heatmap corresponds to its relative significance in predicting its class. A radiologist confirmed that in the case of Anteroposterior (AP) chest radiograph (\Cref{limecxrcovid}), the ill-defined area of ground glass haze in the right lung parenchyma at mid-zone likely represents COVID-19. Similarly, in the Anteroposterior (AP) chest radiograph (\Cref{limecxrpneumonia}), the wedge-shaped area of consolidation in the right lung parenchyma at the upper zone likely represents pneumonia. The radiologist confirmed that the regions (though not all) highlighted by LIME correspond to the affected regions in case of both COVID-19 and Pneumonia. This points to the applicability of COV-ELM in the identification of medical conditions such as pneumonia and COVID-19.

\section{Conclusions} \label{section5}

The current research is focused on the accurate diagnosis of COVID-19 with high sensitivity. This paper evaluates the suitability of ELM for COVID-19 classification due to its faster convergence, better generalization capability, and shorter training time.  A combination of texture (Spatial, GLDM, HOG, AND GLDM) and frequency features (FFT and DWT) extracted from publicly available CXR image repositories are provided as an input to COV-ELM. The proposed COV-ELM model achieved a macro average f1-score of 0.95 and an overall accuracy of 94.74\% in the present three-class classification scenario. The COV-ELM outperforms other competitive machine learning algorithms with a sensitivity of ${0.94\% \pm 0.02}$ at a 95\% confidence interval. For visualization of the results, LIME has been used to highlight the superpixels that contributed to the prediction of a given class. In the LIME generated heatmaps, the higher intensity regions correspond to the clinically evaluated regions. This establishes the clinical relevance  of the features generated by the proposed model. Further, the training time of COV-ELM being quite low, it can be efficiently retrained on newer bigger and diverse datasets. As part of future work, we would like to investigate how segmentation of the relevant lung regions influences the performance of a classification model.

\bibliographystyle{ios1}           
\bibliography{final}        

%

\end{document}